\documentclass[12pt]{article}
\usepackage{graphicx}

\newcommand\pubnumber{Cavendish-HEP-19/15}
\newcommand\pubdate{\today}

\def\institute{$^*$Universit\"at W\"urzburg, Institut f\"ur Theoretische Physik und Astrophysik, \\
97074 W\"urzburg, Germany \\
$^\P$Universit\"at Z\"urich, Physik-Institut, 8057 Z\"urich, Switzerland \\
$^\ddag$University of Cambridge, Cavendish Laboratory, Cambridge CB3 0HE, UK \\
$^\S$Universit\`{a} di Torino e INFN, 10125 Torino, Italy}
\def\Title#1{\begin{center} {\Large #1 } \end{center}}
\def\Author#1{\begin{center}{ \sc #1} \end{center}}
\def\Address#1{\begin{center}{ \it #1} \end{center}}

\newcommand\pubblock{\rightline{\begin{tabular}{l} \pubnumber\\
         \pubdate  \end{tabular}}}
\newenvironment{Abstract}{\begin{quotation}  }{\end{quotation}}
\newenvironment{Presented}{\begin{quotation} \begin{center} 
             PRESENTED AT\end{center}\bigskip 
      \begin{center}\begin{large}}{\end{large}\end{center} \end{quotation}}
\def\Acknowledgements{\bigskip  \bigskip \begin{center} \begin{large}
             \bf ACKNOWLEDGEMENTS \end{large}\end{center}}
\def\beq{\begin{equation}}
\def\eeq#1{\label{#1}\end{equation}}
\def\eeqn{\end{equation}}

\def\beqa{\begin{eqnarray}}
\def\eeqa#1{\label{#1}\end{eqnarray}}
\def\eeqan{\end{eqnarray}}

\let\bar=\overbar

\def\Dslash{\not{\hbox{\kern-4pt $D$}}}
\def\dslash{\not{\hbox{\kern-2pt $\del$}}}

\def\msb{{\bar{\ssstyle M \kern -1pt S}}}

\begin{document}
\begin{titlepage}
\pubblock

\vfill
\Title{NLO QCD + electroweak predictions for off-shell ttH production at the LHC}
\vfill
% \Author{Ansgar Denner, Jean-Nicolas Lang, Mathieu Pellen\support, and Sandro Uccirati}
\Author{Ansgar Denner$^*$, Jean-Nicolas Lang$^\P$, Mathieu Pellen$^\ddag$\footnote{Speaker}, and Sandro Uccirati$^\S$}
\Address{\institute}
\vfill
\begin{Abstract}
In these proceedings, next-to-leading-order (NLO) QCD and electroweak (EW) corrections to ${\rm p}{\rm p} \to {\rm e}^+ \nu_{\rm e} \mu^- \bar \nu_\mu {\rm b} \bar{\rm b} {\rm H}$ at the LHC are presented.
In these computations the top quarks are considered off their mass shell and all non-resonant contributions are included.
The results are presented in the form of fiducial cross sections and differential distributions.
Moreover, two prescriptions to combine QCD and EW corrections are examined.
\end{Abstract}
\vfill
\begin{Presented}
$12^\mathrm{th}$ International Workshop on Top Quark Physics\\
Beijing, China, September 22--27, 2019
\end{Presented}
\vfill
\end{titlepage}
\def\thefootnote{\fnsymbol{footnote}}
\setcounter{footnote}{0}

\section{Introduction}

At the LHC, the production of pairs of top-antitop quarks in association with a Higgs boson is key to probe the Yukawa coupling of the Higgs boson with the heaviest particle in the Standard Model, the top quark.
To that end, solid theoretical predictions are required.
In particular, the NLO QCD corrections for on-shell top quarks are known since more than 15 years \cite{Beenakker:2001rj,Beenakker:2002nc,Reina:2001sf,Dawson:2003zu}.
The electroweak (EW) ones have been computed only few years ago \cite{Frixione:2014qaa,Frixione:2015zaa,Yu:2014cka}.
In addition, these fixed-order computations have been later supplemented by resummed ones \cite{Kulesza:2015vda,Broggio:2015lya,Broggio:2016lfj,Broggio:2019ewu,Kulesza:2020nfh} and by their matching to parton shower \cite{Frederix:2011zi,Garzelli:2011vp,Hartanto:2015uka}.

The first computation with off-shell top quarks has been obtained in Ref.~\cite{Denner:2015yca}.
The partonic process considered is ${\rm p}{\rm p} \to {\rm e}^+ \nu_{\rm e} \mu^- \bar \nu_\mu {\rm b} \bar{\rm b} {\rm H}$ and the computation featured NLO QCD corrections as well as all off-shell and non-resonant contributions.
Later, the NLO EW corrections have been computed and combined with the previous NLO QCD computation \cite{Denner:2016wet}.
These proceedings reproduce the main results of Ref.~\cite{Denner:2016wet}.

\begin{figure}[htb]
\centering
\includegraphics[height=2.2in]{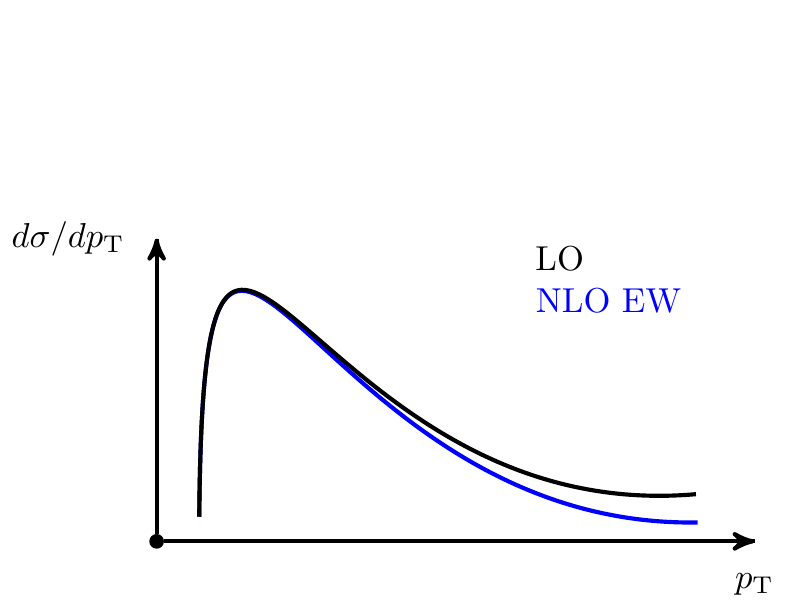}
\includegraphics[height=2.2in]{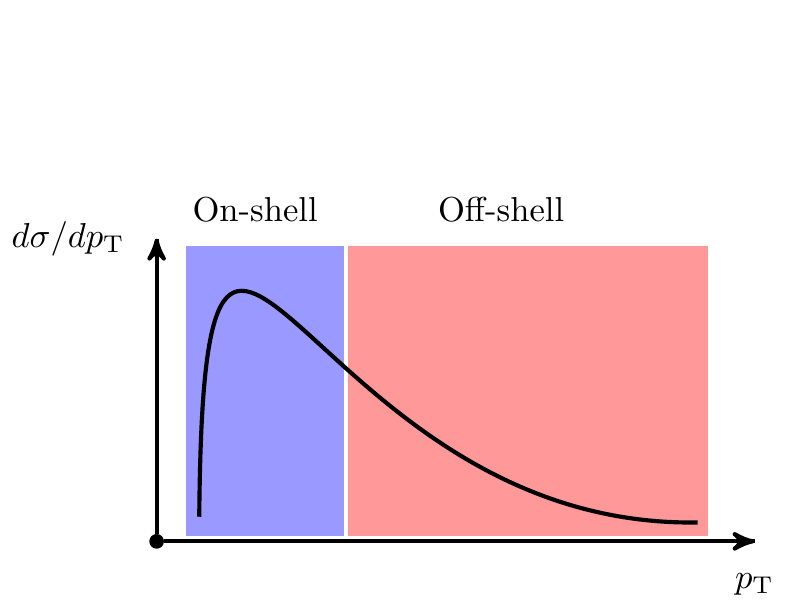}
\caption{Cartoon of the effect of EW corrections in the high-energy tails of distributions (left).
Cartoon of the on-shell region concentrating the bulk of a cross section and the off-shell region in the tails of distributions where non-resonant contributions are sizeable (right).}
\label{fig:cartoon}
\end{figure}

The two main aspects of this computation are the inclusion of EW corrections and of non-resonant contributions.
Typically EW corrections become negatively large at high energy due to the effect of the so-called Sudakov logarithms.
This is illustrated in a cartoon on the left-hand side of Fig.~\ref{fig:cartoon}.
In the same way, non-resonant contributions become sizeable in the tails of distributions.
This is represented on the right-hand side of Fig.~\ref{fig:cartoon} where the on-shell region is dominated by the resonance while the off-shell region features sizeable non-resonant contributions.
In both cases, these effects are critical in the high-energy tails of differential distributions which is the region where new physics is expected to appear.
It is therefore crucial to have a good theoretical description of this part of phase space.
With increasing experimental precision, this region will become accessible in the next few years, making such predictions indispensable.

This contribution is organised as follows:
In the first part, results for the EW corrections are given.
In the second section, the EW corrections are combined with the QCD ones using two different prescriptions.

\section{Electroweak corrections}

In this section we briefly present some differential distributions with EW corrections for ${\rm p}{\rm p} \to {\rm e}^+ \nu_{\rm e} \mu^- \bar \nu_\mu {\rm b} \bar{\rm b} {\rm H}$ at the LHC running at $13{\rm TeV}$.
For the details of the set-up we refer the reader to the original article \cite{Denner:2016wet}.

\begin{figure}[htb]
\centering
\includegraphics[height=2.6in]{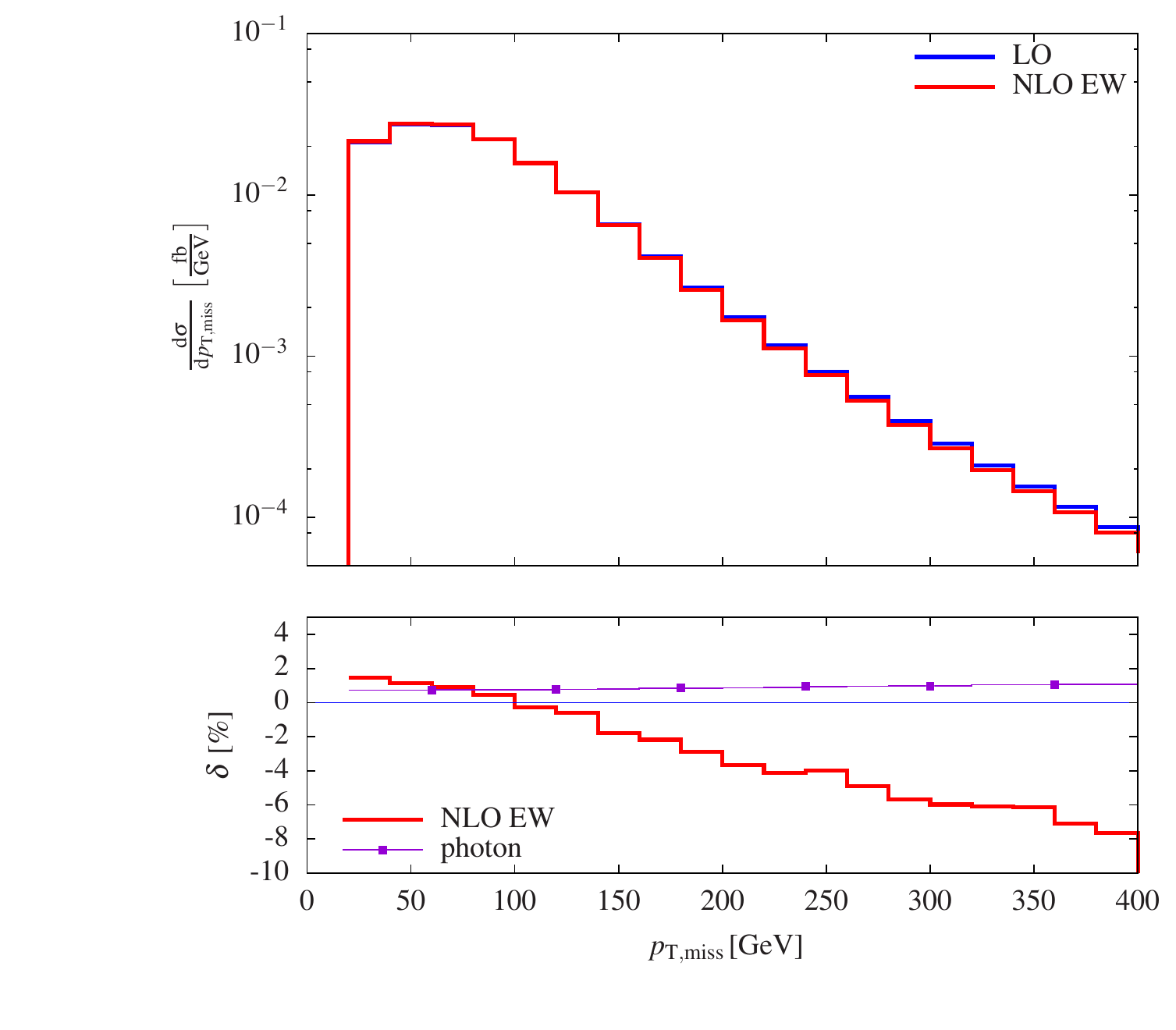}
\includegraphics[height=2.6in]{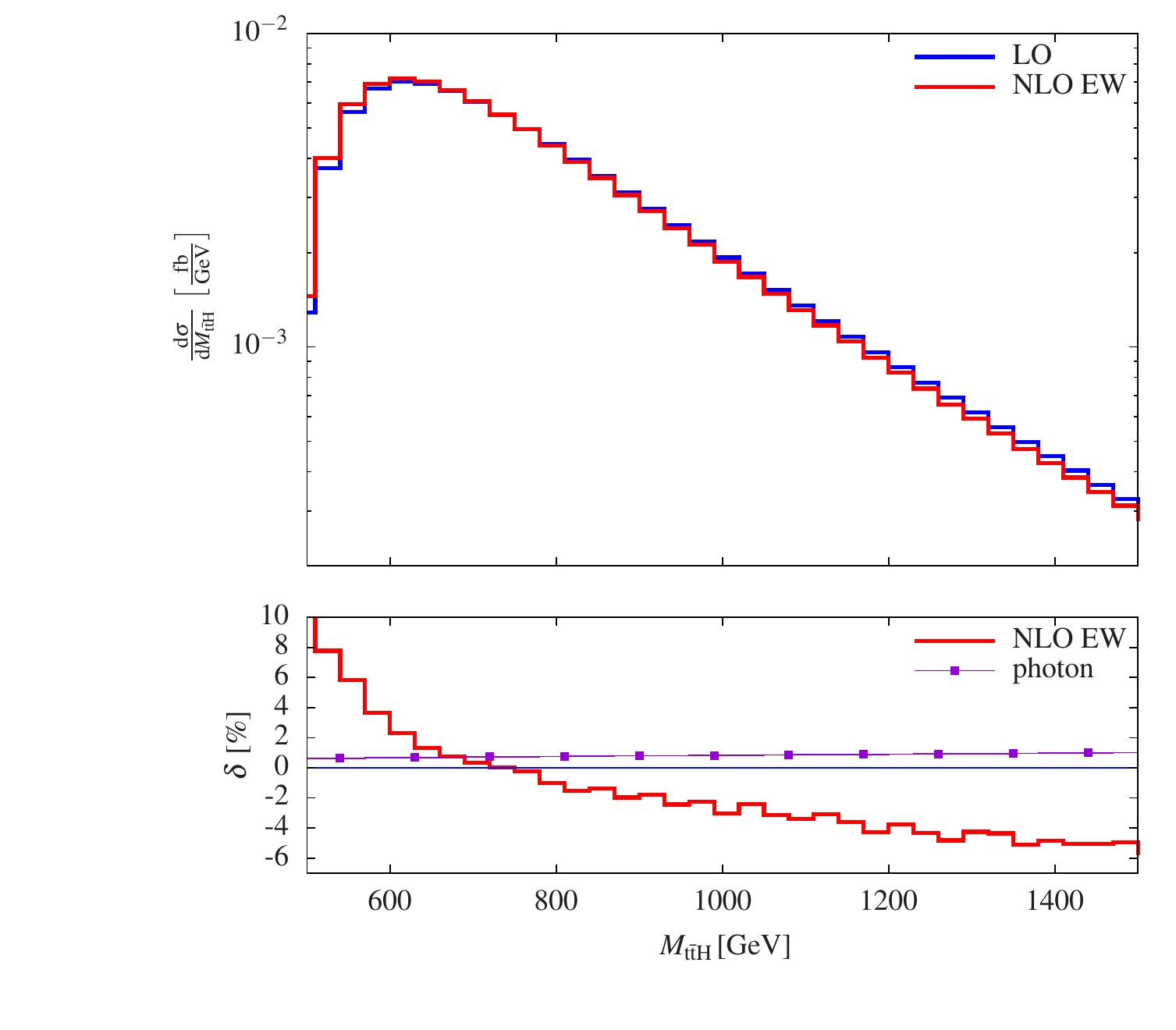}
\caption{Differential distributions for ${\rm p}{\rm p} \to {\rm e}^+ \nu_{\rm e} \mu^- \bar \nu_\mu {\rm b} \bar{\rm b} {\rm H}$ at the LHC running at $13{\rm TeV}$ \cite{Denner:2016wet}:
missing transverse momentum (left) and invariant mass of the ttH system (right).
The plots show the NLO EW corrections as well as the photon-induced contributions.}
\label{fig:ew_dist}
\end{figure}

In Fig.~\ref{fig:ew_dist} (left), the distribution in the missing transverse momentum is shown.
The EW corrections display the typical Sudakov behaviour towards large transverse momentum.
At $400$ GeV, the corrections reach about $-8\%$.
For the invariant mass of the ttH system (right of Fig.~\ref{fig:ew_dist}), the corrections are positively large at threshold (above $10\%$).
They decrease then steadily towards higher invariant mass and are at the level of $-5\%$ around $1.5$ TeV.

The photon-induced contributions shown separately in the plots amount only to a per cent and are uniform across the phase space.
These have been obtained using the LuxQED PDF \cite{Manohar:2016nzj} which feature smaller photon distribution functions than older sets \cite{Denner:2016jyo}.

\section{Combination with QCD corrections}

Let us now turn to the combination of the EW corrections with QCD corrections.
Both NLO corrections are defined as
\begin{equation}
 \sigma^{\mathrm{NLO}}_{\mathrm{QCD}} = \sigma^{\mathrm{Born}} + \delta \sigma^{\mathrm{NLO}}_{\mathrm{QCD}} \qquad {\rm and} \qquad
 \sigma^{\mathrm{NLO}}_{\mathrm{EW}} = \sigma^{\mathrm{Born}} + \delta \sigma^{\mathrm{NLO}}_{\mathrm{EW}} .
\end{equation}

There exist multiple ways of combining them, but we focus here on two prescriptions.
The first one is the so-called \emph{additive} prescription which consists in adding both corrections
\begin{equation}
\sigma^{\mathrm{NLO}}_{\mathrm{QCD+EW}} = \sigma^{\mathrm{Born}} + \delta \sigma^{\mathrm{NLO}}_{\mathrm{QCD}} + \delta \sigma^{\mathrm{NLO}}_{\mathrm{EW}}.
\end{equation}

The second way of combining the two types of corrections is the \emph{multiplicative} way.
It is defined as:
\begin{equation}
 \sigma^{\mathrm{NLO}}_{\mathrm{QCD}\times\mathrm{EW}} = \sigma^{\mathrm{NLO}}_{\mathrm{QCD}} \left( 1 + \frac{\delta \sigma^{\mathrm{NLO}}_{\mathrm{EW}}}{\sigma^{\mathrm{Born}}} \right)
 = \sigma^{\mathrm{NLO}}_{\mathrm{EW}} \left( 1 + \frac{\delta \sigma^{\mathrm{NLO}}_{\mathrm{QCD}}}{\sigma^{\mathrm{Born}}} \right) .
\end{equation}

Both prescriptions are equivalent at NLO accuracy.
Nonetheless, the multiplicative one is usually preferred based on the argument of the factorisation of both types of corrections and consequently gives a better estimate of missing higher-order contributions of mixed type.

In the following, several cross sections and differential distributions are presented for ${\rm p}{\rm p} \to {\rm e}^+ \nu_{\rm e} \mu^- \bar \nu_\mu {\rm b} \bar{\rm b} {\rm H}$ at the LHC running at $13{\rm TeV}$.
Again, we refer the interested reader to Ref.~\cite{Denner:2016wet} for details.

In Table~\ref{tab:crosssection}, LO, NLO QCD, and NLO EW predictions are presented.\footnote{Note that $\sigma^{\rm LO}$ and $\sigma^{\rm Born}$ are not identical because different top widths have been used (see Ref.~\cite{Denner:2016wet}).}
In addition, the numerical predictions for the combinations of the two NLO corrections are given.
As can be seen, the difference between the two prescriptions is irrelevant.
This is simply due to the fact that the NLO EW corrections are very small at the level of the total cross section.

\begin{table}[t]
\begin{center}
\begin{tabular}{cccccc}  
 $\sigma^{\rm LO}$ & $\sigma^{\rm Born}$ & $\sigma^{\mathrm{NLO}}_{\mathrm{QCD}}$ & $\sigma^{\mathrm{NLO}}_{\mathrm{EW}}$ & $\sigma^{\mathrm{NLO}}_{\mathrm{QCD+EW}}$  & $\sigma^{\mathrm{NLO}}_{\mathrm{QCD}\times\mathrm{EW}}$ \\
  \hline\hline
$ 2.4817(1) $ &  $ 2.7815(1) $ & $ 2.866(1) $ & $ 2.721(3) $ & $ 2.806 $ & $ 2.804 $ \\
  \hline
\end{tabular}
\caption{Cross sections in femto barn for ${\rm p}{\rm p} \to {\rm e}^+ \nu_{\rm e} \mu^- \bar \nu_\mu {\rm b} \bar{\rm b} {\rm H}$ at LO, NLO QCD, and NLO EW \cite{Denner:2016wet}.
The \emph{Born} value corresponds to the tree-level contribution entering the NLO predictions.
The two most right columns correspond to the two prescriptions to combine EW and QCD corrections.
The digits in parenthesis represent the numerical error.}
\label{tab:crosssection}
\end{center}
\end{table}

\begin{figure}[htb]
\centering
\includegraphics[height=2.6in]{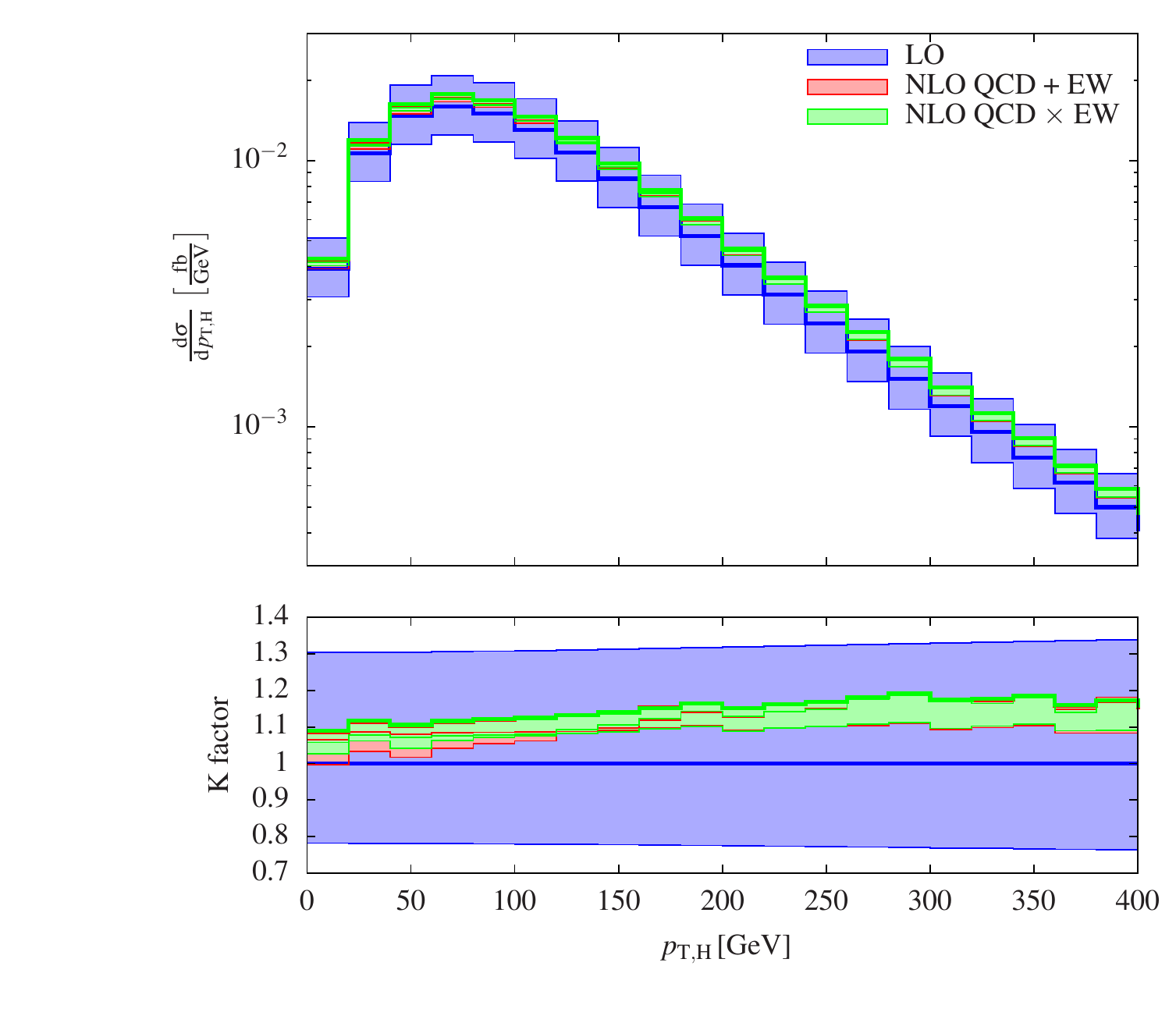}
\includegraphics[height=2.6in]{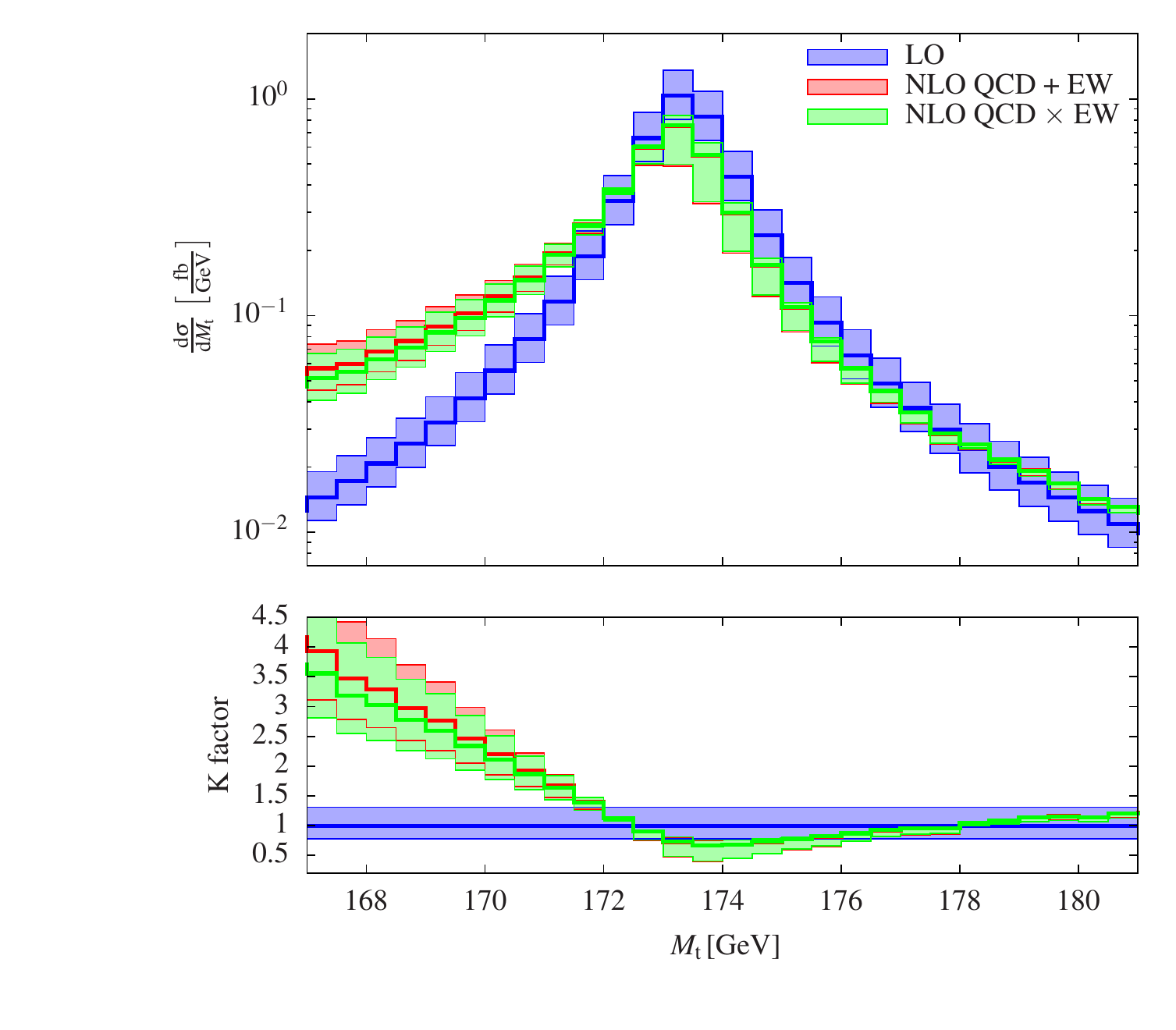}
\caption{Differential distributions for ${\rm p}{\rm p} \to {\rm e}^+ \nu_{\rm e} \mu^- \bar \nu_\mu {\rm b} \bar{\rm b} {\rm H}$ at the LHC running at $13{\rm TeV}$ \cite{Denner:2016wet}:
transverse momentum of the Higgs boson (left) and invariant mass of the reconstructed top quark (right).
Both the additive and the multiplicative combination of the NLO EW and QCD corrections are shown.}
\label{fig:ew_dist2}
\end{figure}

In Fig.~\ref{fig:ew_dist2}, two differential distributions are shown for both the additive and the multiplicative combination of the NLO EW and QCD corrections.
The plot on the left shows the transverse momentum of the Higgs boson.
As for the total cross section, both prescriptions are very similar.
This originates from the fact that the total corrections of this distribution are dominated by the QCD ones, while the EW corrections are small making the two combinations effectively identical.
Turning to the differential distribution in the invariant mass of the reconstructed top quark (right), one can actually see differences between the two prescriptions.
This is due to the large radiative tail for both types of corrections below the top-quark resonance, which leads to significant differences between the two combinations.
These differences can give an estimate of missing higher-order corrections of mixed type.

\section{Conclusion}

In these proceedings results for the electroweak corrections to ${\rm p}{\rm p} \to {\rm e}^+ \nu_{\rm e} \mu^- \bar \nu_\mu {\rm b} \bar{\rm b} {\rm H}$ at the LHC have been presented.
The main advantage of this computation is that it features EW corrections together with all off-shell and non-resonant effects.
These corrections have been combined with the QCD ones following two prescriptions.
All these effects will become particularly relevant in the next few years when the experimental precision will allow to explore further high-energy tails of differential distributions.

\Acknowledgements

We are grateful to Robert Feger for providing and supporting the code MoCaNLO. The work of A.D. and M.P. was supported by the Bundesministerium f\"ur Bildung und Forschung (BMBF) under contract no. 05H15WWCA1.
The research of M.P. has received funding from the European Research Council (ERC) under the European Union's Horizon 2020 research and innovation programme (grant agreement No 683211).

\end{document}